\documentclass[journal=jacsat,manuscript=article, layout=traditional]{achemso}

\setkeys{acs}{maxauthors = 0}
\usepackage[version=3]{mhchem} 
\usepackage[LGRgreek]{mathastext}
\usepackage[percent]{overpic}
\usepackage{upgreek}
\usepackage{amsmath}
\usepackage{xcolor}
\usepackage{comment}

\author{Marko Mladenovi\'c}
\email{mmladenovic@iis.ee.ethz.ch}
\affiliation[ETH Zürich]
{Integrated Systems Laboratory, Department of Information Technology and Electrical Engineering, ETH Zurich, CH-8092 Zurich, Switzerland}

\author{Manasa Kaniselvan}
\affiliation[ETH Zürich]
{Integrated Systems Laboratory, Department of Information Technology and Electrical Engineering, ETH Zurich, CH-8092 Zurich, Switzerland}

\author{Christoph Weilenmann}
\affiliation[ETH Zürich]
{Integrated Systems Laboratory, Department of Information Technology and Electrical Engineering, ETH Zurich, CH-8092 Zurich, Switzerland}

\author{Alexandros Emboras}
\affiliation[ETH Zürich]
{Integrated Systems Laboratory, Department of Information Technology and Electrical Engineering, ETH Zurich, CH-8092 Zurich, Switzerland}

\author{Mathieu Luisier}
\affiliation[ETH Zürich]
{Integrated Systems Laboratory, Department of Information Technology and Electrical Engineering, ETH Zurich, CH-8092 Zurich, Switzerland}

\title
  {Termination-Dependent Resistive Switching in SrTiO$_3$ Valence Change Memory Cells}

\begin{document}


\begin{abstract}

Valence change memory (VCM) cells based on SrTiO$_3$ (STO), a perovskite oxide, are a promising type of emerging memory device. While the operational principle of most VCM cells relies on the growth and dissolution of one or multiple conductive filaments, those based on STO are known to exhibit a distinctive, `interface-type' switching, which is associated with the modulation of the Schottky barrier at their active electrode. Still, a detailed picture of the processes that lead to interface-type switching is not available. In this work, we use a fully atomistic and \textit{ab initio} model to study the resistive switching of a Pt-STO-Ti stack. We identify that the termination of the crystalline STO plays a decisive role in the switching mechanism, depending on the relative band alignment between the material and the Pt electrode. In particular, we show that the accumulation of oxygen vacancies at the Pt side can be at the origin of resistive switching in TiO$_2$-terminated devices by lowering the conduction band minimum of the STO layer, thus facilitating transmission through the Schottky barrier. Moreover, we investigate the possibility of filamentary switching in STO and reveal that it is most likely to occur at the Pt electrode of the SrO-terminated cells.
  
\end{abstract}


\newpage

\section{Introduction}
Valence change memory (VCM) cells, sometimes referred to as memristors, are resistive switching devices that have been proposed as non-volatile storage units in future hardware architectures\cite{VCM_2, VCM_3, VCM_5} and as solid-state synapses in neuromorphic computing applications\cite{VCM_1, VCM_8}. Their advantage over other memristive systems lies in their scalability, easy production, and large dynamic range, which corresponds to the ratio between their high (HRS) and low resistance states (LRS). Most VCM cells achieve resistive switching between their HRS and LRS through filamentary soft dielectric breakdown within an oxide layer. These filaments are conductive, typically consist of oxygen vacancies, and can be reversibly ruptured (RESET process) and restored (SET process). Such behavior typically leads to abrupt SET processes and linear current-voltage (I-V) characteristics in the LRS state. \cite{VCM_1, VCM_2, VCM_3, VCM_4, VCM_5, VCM_6, VCM_8, VCM_9}.

Contrary to this abrupt switching behavior displayed by majority of oxide-based devices, VCM cells using Strontium titanate (SrTiO$_3$, shorter STO), a perovskite oxide, are reported to display more gradual conductance changes and distinctively nonlinear I-V characteristics \cite{STO_6}. The ability to achieve this graduate conductance modulation makes them ideal for analog option, as required for electronic synapses \cite{Christoph}. The switching mechanism responsible for this phenomenon is assumed to be non-filamentary and localized at contact-oxide interfaces, as supported by imaging of conductive species in these regions \cite{STO_6, STO_7, STO_10}. It is thus generally attributed to a reversible modulation of the Schottky barrier height at a high work function contact \cite{STO_1, STO_6, STO_7}. However, the physical process leading to this modulation remains unclear. Three main mechanisms have been so far proposed in the literature: (i) accumulation and migration of oxygen vacancies \cite{STO_2, STO_3, STO_11, STO_7}, (ii) charge trapping/de-trapping at the interface \cite{STO_5, STO_4, STO_7}, and (iii) oxygen vacancy-ion pair generation \cite{STO_6, STO_10}. It could be that several of these mechanisms occur simultaneously \cite{STO_7}. Previous reports also suggested that a conductive filament might be present in STO devices\cite{STO_8, STO_10, STO_6}, potentially resulting in the presence of both filamentary- and interface-type switching.

These two switching mechanisms, filamentary- and interface-type, do not only differ from each other with respect to the linearity of their I-V curves, but may also differ in terms of their switching direction. The selection of one mechanism over the other depends on the voltage amplitude of the first HRS-to-LRS transition and on the sweep rate \cite{STO_9, STO_10}. However, there is a general consensus that the counter-eightwise switching (where the transition from LRS to HRS happens when a positive voltage is applied to the active electrode) can be associated with oxygen vacancy migration, while the opposite switching direction (the transition from HRS to LRS happens when a positive voltage is applied to the active electrode) can be associated with charge trapping/de-trapping or oxygen vacancy-ion pair generation at the interface \cite{STO_9, STO_10, STO_6, STO_11}. The rich physics of STO-based VCM cells and its dependence on multiple parameters incentivize thorough theoretical investigations capable of providing insights into the functionality of these devices and of revealing the nature of their switching mechanism(s). 

Through a multiscale approach combining density functional theory (DFT), kinetic Monte Carlo (KMC), and \textit{ab initio} quantum transport calculations, we propose a detailed study on resistive switching in a VCM cell made of an STO oxide surrounded by a Pt and Ti electrode, forming a Schottky and Ohmic contact, respectively. We initially construct so-called static models with a pre-defined distribution of oxygen vacancies (V$_O$) to test different possibilities of resistive switching through vacancy migration, one of the proposed underlying mechanisms. Based on the results of these models, we run KMC simulations to obtain more realistic V$_O$ distributions and to monitor full switching cycles for both interface- and filamentary-type switching. We find that the edge-termination of the SrTiO$_3$ oxide at the electrode interfaces has a strong influence on the resulting metal-oxide barrier heights. It determines the nature of the electronic current flowing through the oxide and thus the resulting resistive switching type. Notably, in a TiO$_2$-terminated STO layer, the modulation of the Schottky barrier at the Pt electrode by oxygen vacancy migration and accumulation leads to non-volatile resistive switching. On the contrary, a SrO-terminated insulator layer does not clearly show this effect, and is more likely to exhibit filamentary-type switching. 
Our results indicate that engineering the termination of SrTiO$_3$ VCM cells could be a viable approach to controlling their switching dynamics. 

This paper is organized as follows: We first introduce the signatures of filamentary and interface-type switching, using experimental data from a similar STO stack. We then simulate the electronic properties and current flow through devices with predefined `static' arrangements of vacancies, designed to mimic the interface-type and filamentary regimes. Here we explore the effect of the two different interface terminations that the crystalline STO layer can exhibit, and find that this has a significant influence on the extent to which interface-type switching can be achieved. Based on these findings, we move to a dynamic model of both interface- and filamentary-type switching, from which we can reproduce the characteristic switching direction of both regimes. Along the way, we introduce the computational approaches used for each of these studies. 

\section{Computational methods}

All electronic structure calculations are performed with density functional theory (DFT) as implemented in the CP2K code and its Gaussian-type orbitals (GTOs) \cite{CP2K}, while kinetic Monte Carlo (KMC)\cite{KMC} is used to model structural changes under applied electric field. The electronic current flowing through the constructed VCM cells is computed via the Quantum Transmitting Boundary Method (QTBM) \cite{OMEN}. Prior to assembling Pt-STO-Ti pristine devices, bulk structures for the Ti and Pt electrodes as well as for crystalline SrTiO$_3$ are created. These three blocks are then attached together and the distance between each of them is optimized (details behind the distance optimization are given in Fig.\;S1 of the Supporting Information). A strain of 0.7 $\%$ is applied to the Pt electrode along both the \textit{y} and \textit{z} directions, which define the device cross section, \textit{x} being the axis connecting the metallic electrodes. Conversely, the Ti electrode undergoes a strain of 0.2 $\%$ and 2.9 $\%$ along these axes. 

 Structural relaxations are performed using the L-BFGS minimization method of CP2K with a double-$\zeta$ polarization (DZVP) basis set\cite{dzvp} and the PBE functional, based on the Generalized Gradient Approximation\cite{pbe}. Convergence criteria of 4.5 $\times$ $10^{-4}$ Ha/Bohr for forces and 3 $\times$ $10^{-3}$ Bohr for the geometry change are used. The plane-wave cutoff is set to 500 Ry, while a cutoff of 60 Ry is employed to map the GTOs onto the plane-wave grid. KMC simulations of the switching process require the activation energy of events such as oxygen vacancy diffusion or vacancy-ion pair generation as input parameters. We calculate these quantities with the Nudged Elastic Band (NEB) method\cite{NEB}, with a total of 7 intermediate images per process. Due to the cubic symmetry of SrTiO$_3$, these quantities are calculated for a single path. More details on the KMC simulations and activation energies are given in the Supporting Information. To generate the Hamiltonian and overlap matrices that serve as inputs to transport calculations, a single-$\zeta$ polarization basis set is used to minimize the computational complexity. The Hamiltonian elements smaller than 10$^{-6}$ Ha are not considered in the transport calculations. Still, the band gap underestimation of DFT is corrected through the Hubbard (DFT+U) approach \cite{DFT_U} with parameters U - J = 9 eV applied to the 3d orbitals of Ti (details behind the Hubbard parameter identification are given in Fig.\;S2). 
Open boundary conditions are applied along the transport direction (\textit{x}-axis connecting the electrodes in \textbf{Fig.~\ref{fig:structure}}) to allow for electron injection and to determine the transmission function through the devices in the coherent limit of transport. 

\begin{figure}[!t]
\centering\includegraphics[width=\columnwidth]{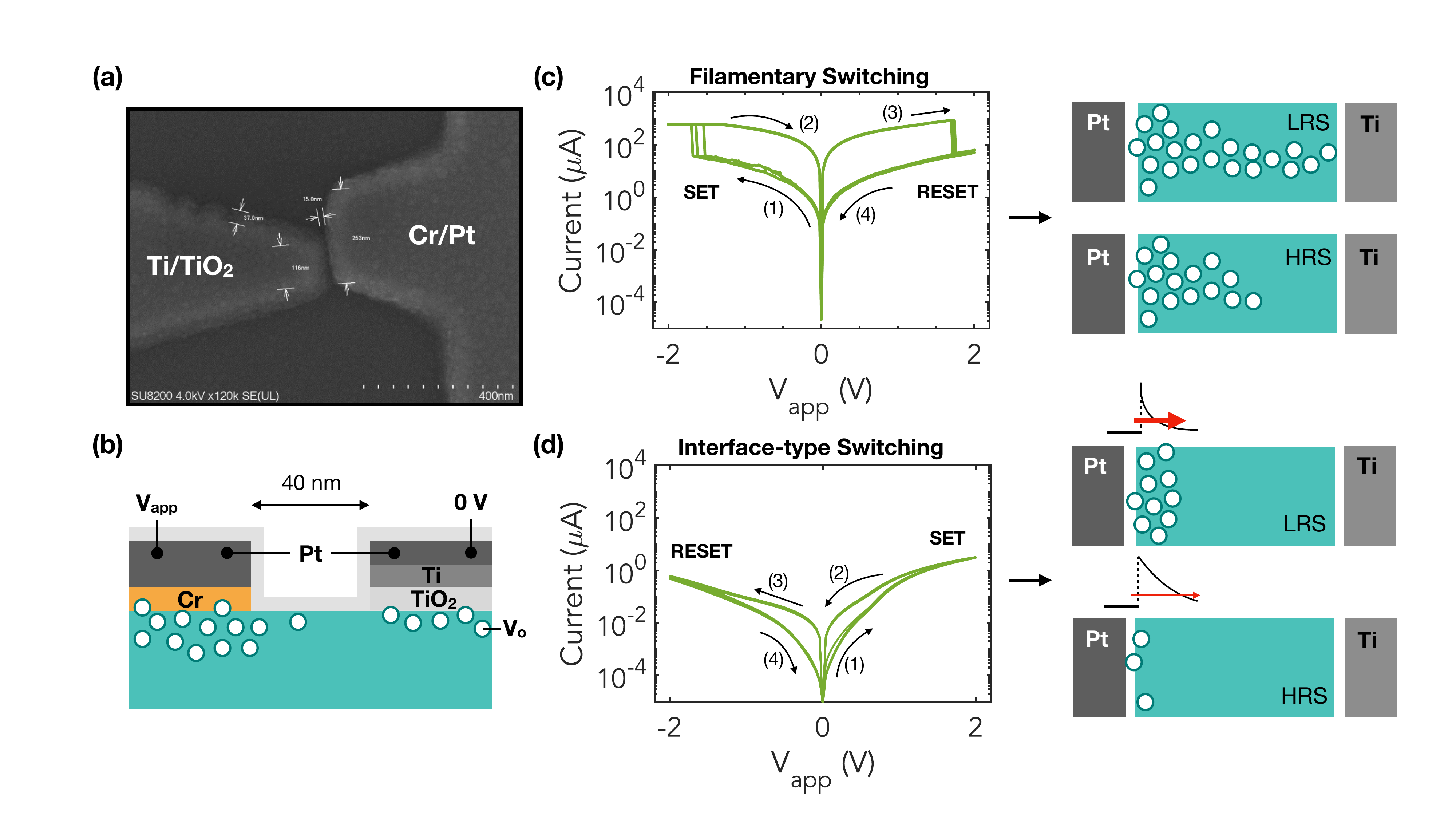}
\caption{(a) TEM image and (b) layout of the fabricated Pt-SrTiO$_3$-Ti-Pt stack. The Pt/Ti electrode is grounded, while a voltage V$_{app}$ is applied to the Pt one. Experimentally measured I-V characteristics of (c) filamentary- and (d) interface-type switching. The sub-plots on the right illustrate the underlying oxygen vacancy distributions (white circle) associated with both switching mechanisms. The red arrows indicate the flow of the electrons through the Schottky barrier in the case of interface-type switching and its magnitude in the HRS and LRS, respectively.}
\label{fig:exp1}
\end{figure}

\begin{figure}[!t]
\centering\includegraphics[width=\columnwidth]{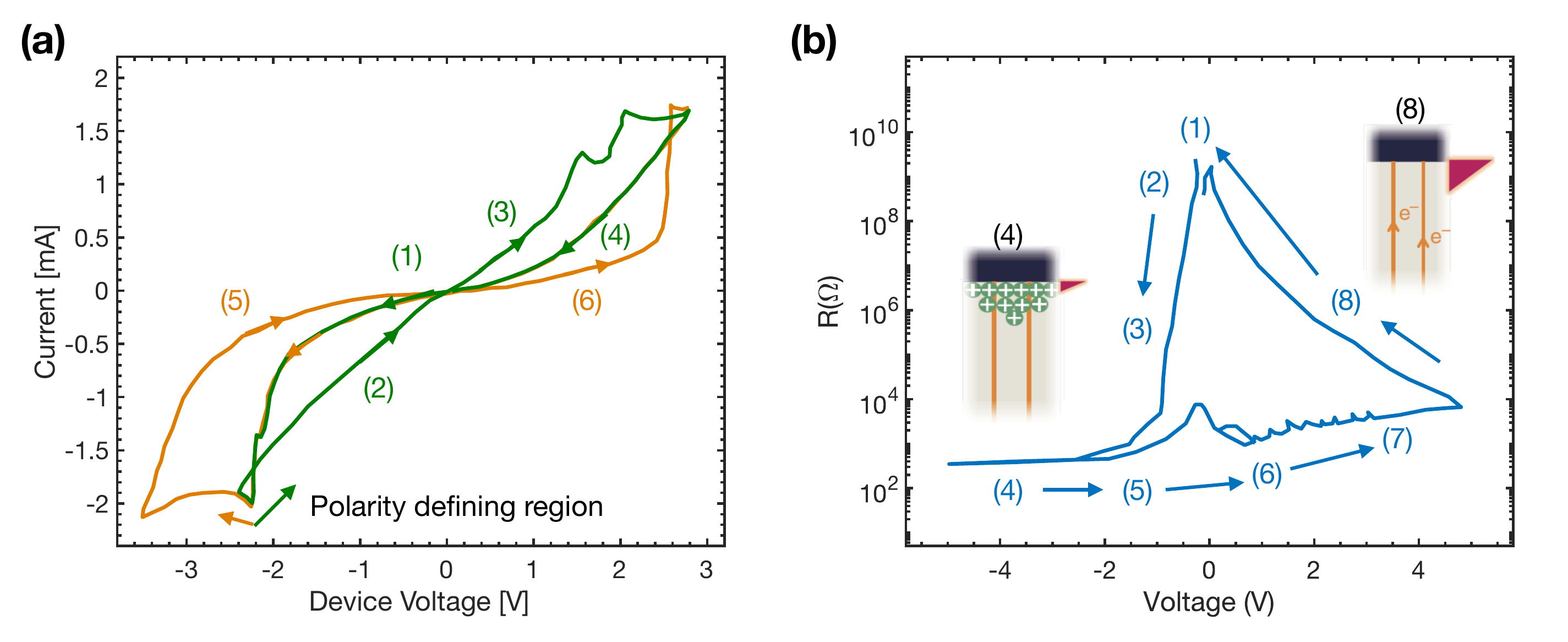}
\caption{(a) Counter-eightwise (green) and eightwise (orange) resistive switching in Fe-doped SrTiO$_3$ device. (b) Resistive switching in Sm-doped CeO$_2$: SrTiO$_3$ mixture due to oxygen vacancy migration. The insets show the assumed LRS (left) and HRS (right) states. Sub-plot (a) is based on the data from [Muenstermann, R, et al., 2010, Wiley], sub-plot (b) from [Cho, S., et al., 2016, Nature Portfolio].}
\label{fig:exp2}
\end{figure}

\begin{figure}[!t]
\centering\includegraphics[width=\columnwidth]{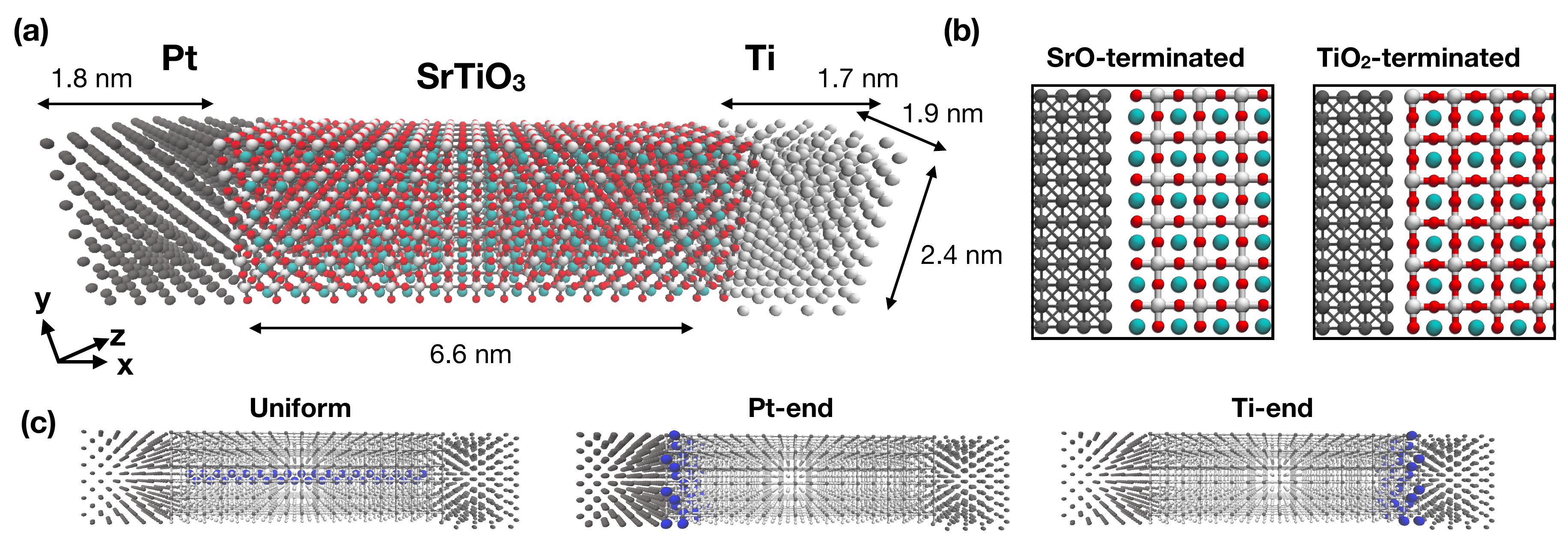}
\caption{(a) Typical device structure consisting of a Pt-SrTiO$_3$-Ti stack with dimensions indicated. (b) Visualizations of the SrO (left) and the TiO$_2$ (right) terminations. Sr atoms are shown in cyan, Ti in white, O in red, and Pt in gray. (c) Different distributions of oxygen vacancies (uniform, grouped at the Pt end or at the Ti end). These oxygen vacancies are represented as dark blue spheres.
}
\label{fig:structure}
\end{figure}

\section{Results}

As an initial step, we consider experimental measurements on a Pt-SrTiO$_3$-Ti device. A Transmission Electron Microscopy (TEM) image of the fabricated stack is shown in \textbf{Fig.~\ref{fig:exp1}(a)}, while the corresponding layout is depicted in \textbf{Fig.~\ref{fig:exp1}(b)}. The reported ``current-voltage'' (I-V) measurements are conducted with (\textbf{Fig.~\ref{fig:exp1}(c), left}) and without (\textbf{Fig.~\ref{fig:exp1}(d), right}) an initial forming process, followed by voltage sweeps between -2 and 2 V in both cases. The forming step induces a filament made of oxygen vacancies so that the I-V characteristics exhibit a counter-eightwise, abrupt resistive switching, typical of filamentary-type cells (\textbf{Fig.~\ref{fig:exp1}(c)}). In this switching type, creation (SET) and rupture (RESET) of the filament, accompanied with lengthening and shortening of the conductive path, is the dominant mechanism. On the contrary, the I-V characteristics of the VCM cells without the initial forming process exhibit a non-linear eightwise switching, which is the signature of interface-type switching (\textbf{Fig.~\ref{fig:exp1}(d)}). This switching type assumes a modulation of the Schottky barrier at the active electrode, whose lowering and shortening leads to a transition from LRS to HRS (\textbf{Fig.~\ref{fig:exp1}(d), right}). Modulation of the Schottky barrier, on the other hand, can be achieved by tuning the amount of electrostatic charge at the interface, for example through oxygen vacancy accumulation. When comparing the two switching types, we find that the current is roughly two orders of magnitude higher in the filamentary case. Details on device fabrication process and on experimental measurements are available in Ref.~\cite{Christoph}.

As mentioned before, the different switching directions reported for the interface-type switching are attributed to different underlying mechanisms, namely vacancy migration for the counter-eightwise, and charge trapping/de-trapping and/or vacancy generation for the eightwise switching direction. The transition from one regime to the other can even be achieved without ``playing'' with the forming step, for example by applying a larger negative voltage to the active electrode of a Pt-Fe: SrTiO$_3$-Nb: SrTiO$_3$ device (\textbf{Fig.~\ref{fig:exp2}(a)}) \cite{STO_10}. Namely, the operation mode of the device can be converted from the counter-eightwise (1-2-3-4) to the eightwise one (1-5-6-4) if instead of the SET process at around -2.2 V, the voltage is increased up to -3.5 V which leads to a RESET to an even lower HRS state. Therefore, two or more switching mechanisms may co-exist in a single device. The possibility of resistive switching by oxygen vacancy migration is further corroborated by findings in Ref.~\cite{STO_2}, where the LRS in Sm-doped Pt-CeO$_2$: SrTiO$_3$-Nb: SrTiO$_3$ device is attributed to the accumulation of oxygen vacancies at the Pt electrode (\textbf{Fig.~\ref{fig:exp2}(b)}). For these reasons, in the rest of the paper, we primarily focus on modeling the counter-eightwise switching by vacancy migration.

\subsection{Static model of resistive switching in SrTiO$_3$}

Crystalline SrTiO$_3$ consists of alternating planes of SrO and TiO$_2$. The nature of the plane situated at the interface with the metallic electrodes alters the physics of the formed contact \cite{Z3}. Therefore, we investigate here the influence of this interface termination on the transport characteristics of SrTiO$_3$ VCM cells, by constructing Pt-SrTiO$_3$-Ti stacks (\textbf{Fig.~\ref{fig:structure}(a)}) with SrO- and TiO$_2$-terminated interfaces on both sides (Pt and Ti), as illustrated in \textbf{Fig.~\ref{fig:structure}(b) left and right}, respectively. As first test, we investigate the possibility of switching the Pt-SrTiO$_3$-Ti stack between its LRS and HRS states solely by considering vacancy migration. The latter mechanism was proposed for non-filamentary resistive switching in SrTiO$_3$\cite{STO_2}. To achieve that, we manually distribute oxygen vacancies throughout the STO layer by removing oxygen atoms from the lattice, either uniformly or at one of the contact-oxide interfaces, as shown in \textbf{Fig.~\ref{fig:structure}(c)}. The resulting configurations represent idealized filamentary- and interface-type LRS and HRS states. 

\begin{figure}
\centering\includegraphics[width=\columnwidth]{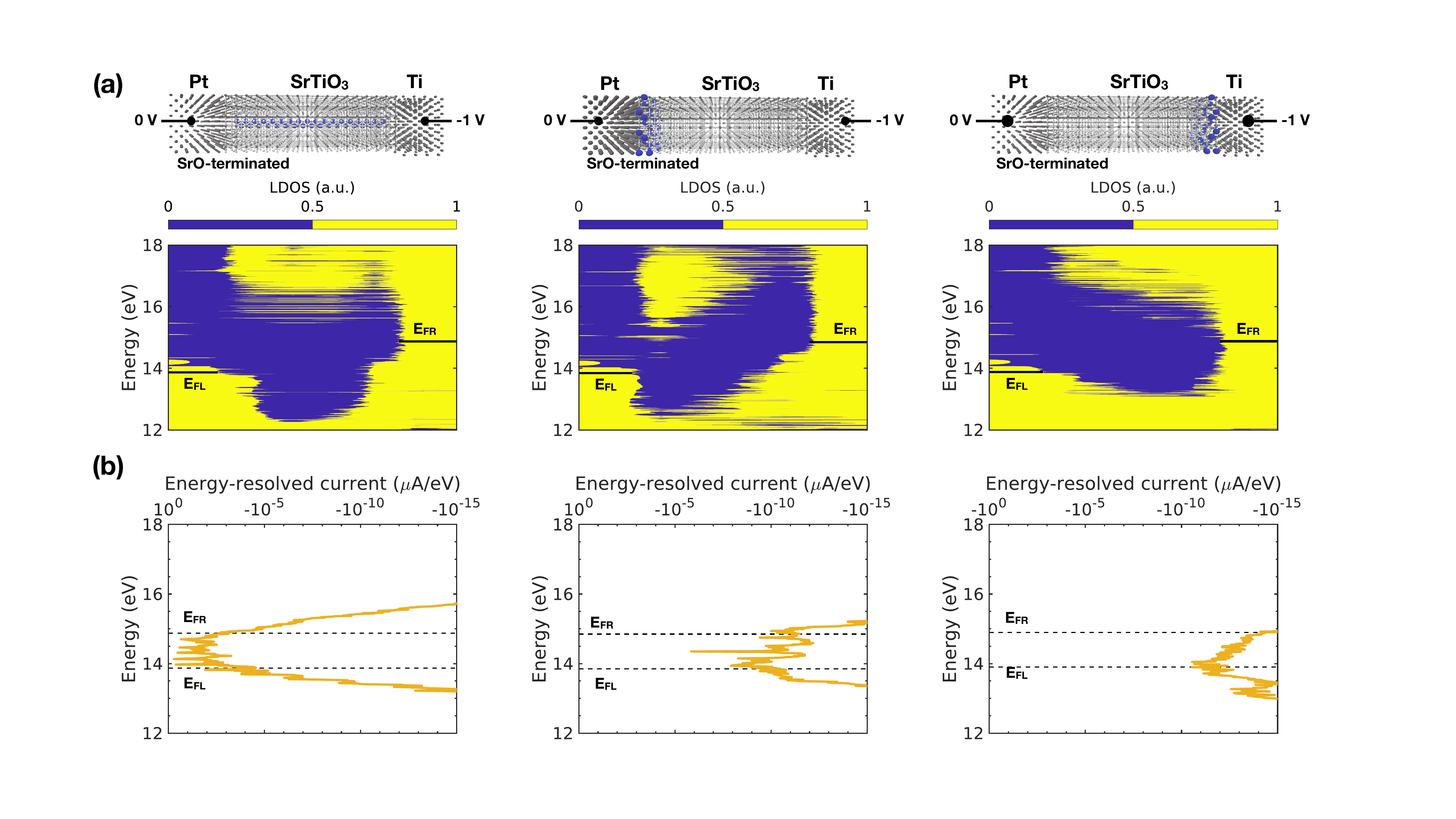}
\caption{(a) Local density-of-states (LDOS) and (b) energy-resolved electronic current extracted from SrO-terminated Pt-SrTiO$_3$-Ti stacks under an applied bias of -1 V to the Ti electrode (Pt electrode is grounded). The atomic configurations with a uniform distribution of the oxygen vacancies (left), V$_O$ accumulated at the Pt electrode (middle), and V$_O$ accumulated at the Ti electrode (right) are displayed. They correspond to the scenarios defined in \textbf{Fig.~\ref{fig:structure}(c)}. Yellow denotes a high DOS, while blue refers to a low DOS. The positions of the left ($E_{FL}$) and right ($E_{FR}$) Fermi levels are indicated in LDOS and current plots.
}
\label{fig:ldos_sro}
\end{figure}

We analyze the influence of the oxygen vacancy distribution and interface-termination by inspecting the local density-of-states (LDOS) and energy-resolved currents of all possible configurations, under the same bias conditions, starting in \textbf{Fig.~\ref{fig:ldos_sro}} with SrO-terminated SrTiO$_3$ devices. We observe that the current is the highest when the vacancies are distributed uniformly, forming an ideal filament, compared to the model where they accumulate at either the Pt or Ti end. This implies that hopping through defect states\cite{hopping} created in the SrTiO$_3$ band gap is the dominant transport mechanism: The electrical current magnitude is indeed by far the largest at energies in between the conduction and valence band edges of STO. Accumulation of vacancies at one of the oxide-metal interfaces, meanwhile, has a minimal impact on the tunneling probability from one electrode to the other. Therefore, we conclude that the modulation of the Schottky barrier height via vacancy accumulation/depletion cannot induce resistive switching in purely SrO-terminated SrTiO$_3$ VCM cells, and filamentary-type switching is more likely to occur.

\begin{figure}
\centering
\includegraphics[width=\columnwidth]{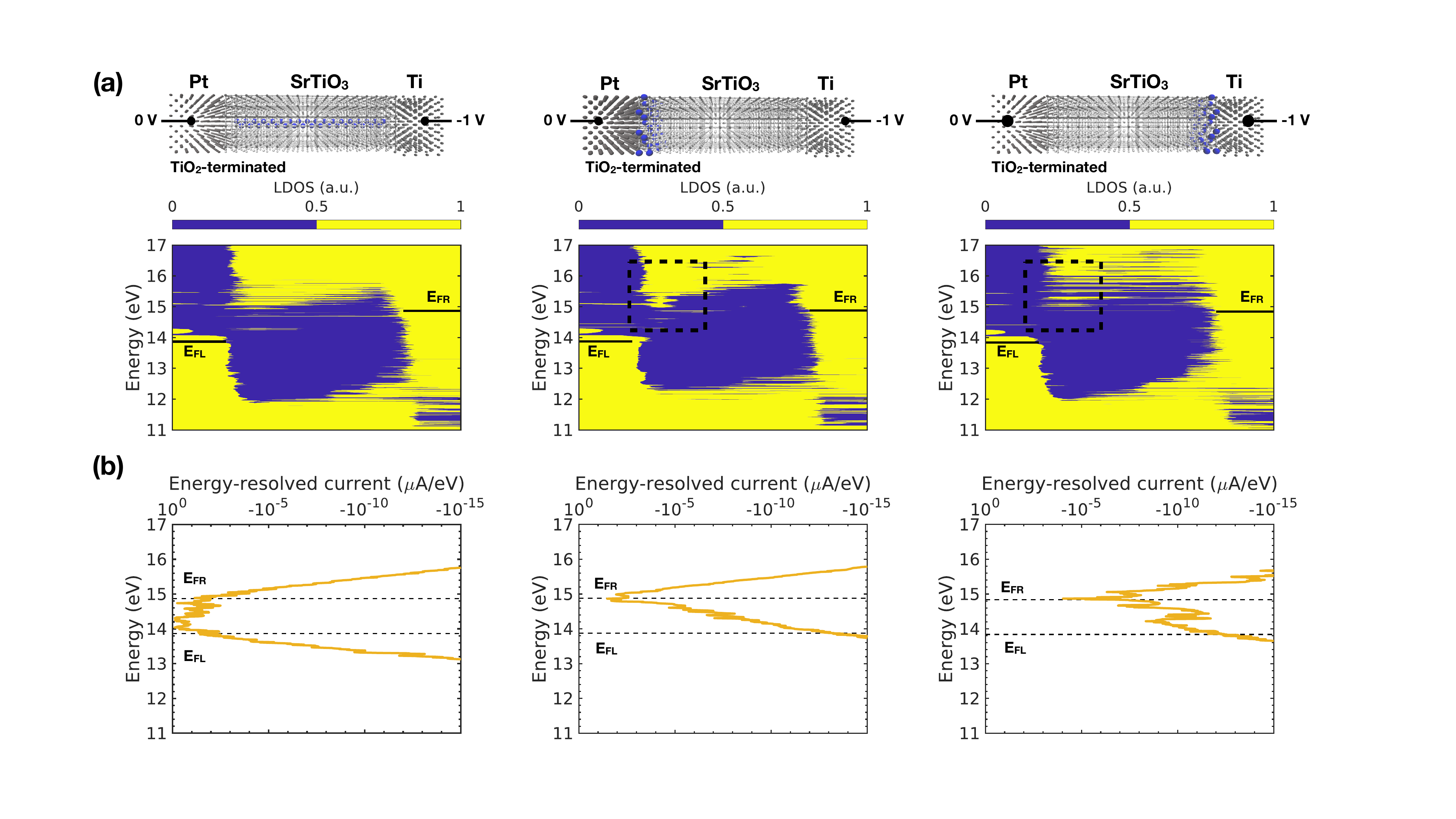}
\caption{Same as \textbf{Fig.~\ref{fig:ldos_sro}} but for TiO$_2$-terminated Pt-SrTiO$_3$-Ti. The dashed frames in the middle and right sub-plots in (a) highlight the differences in LDOS around the Schottky barrier. }
\label{fig:ldos_tio2}
\end{figure}

The situation is different in TiO$_2$-terminated SrTiO$_3$ structures (\textbf{Fig.~\ref{fig:ldos_tio2}}) as the barrier at the Ti interface is greatly reduced. Electrical current is thus primarily injected into the conduction band of the oxide, rather than into band gap states. The height of the Schottky barrier at the Pt-end (left electrode) can be modulated by locally accumulating oxygen vacancies. Doing so lowers the conduction band minimum of SrTiO$_3$, which decreases the Pt-SrTiO$_3$ band offset and thus facilitates transmission through the reduced Schottky barrier height. The resulting configuration can be interpreted as a possible LRS, the HRS corresponding to the case where the vacancies are pushed towards the Ti electrode on the right. Once this happens, the Schottky barrier returns to its original state, the electrical current diminishes, and the device is no more conductive. Note, however, that the highest current is still obtained for the uniform distribution of vacancies. This is consistent with the experimental findings that the electro-formed devices exhibit larger conductance values than the non-filamentary ones (\textbf{Fig.~\ref{fig:exp1}(c)} and \textbf{(d)}). Additionally, we assumed that the main interface governing electric transport is at the Pt end, even though our calculations do not produce an Ohmic contact at the Ti end, as expected experimentally\cite{STO_1}. A small Schottky barrier is present there as well. Nonetheless, simulations of the TiO$_2$-terminated SrTiO$_3$ cell clearly demonstrate that interface-type switching based on the migration of oxygen vacancies from one electrode to the other can occur in such devices and that a formed filament is not necessary to define two resistance states. 

\begin{figure}
\centering
\includegraphics[width=\textwidth]{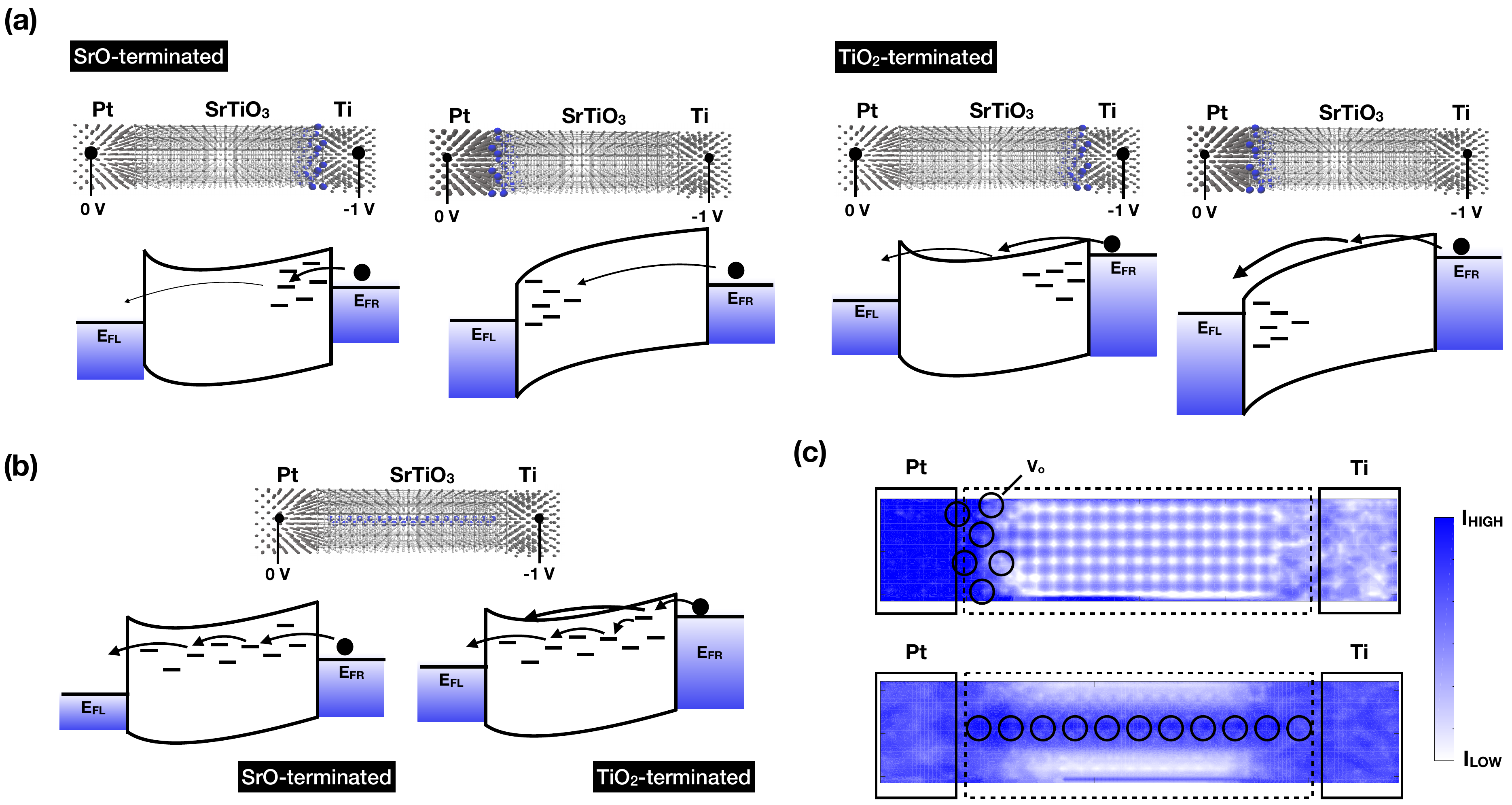}
\caption{Schematic view of the band diagrams corresponding to (a) vacancies distributed at either the Ti or Pt end and (b) uniform distribution of vacancies for both SrO-(left) and TiO$_2$-(right) terminated SrTiO$_3$ cells. 
The conduction and valence band edges, as well as the defect energy levels were qualitatively derived from the LDOS plots shown in \textbf{Figs.~\ref{fig:ldos_sro}} and \textbf{\ref{fig:ldos_tio2}}. (c) Spatial distribution of the electrical current corresponding to the LRS of the TiO$_2$-(top) and the SrO- terminated (bottom) device where interface- and filamentary-type switching dominates, respectively. 
}
\label{fig:current}
\end{figure}

We summarize the above-discussed results in schematic diagrams in \textbf{Fig.~\ref{fig:current}}. From this first set of calculations, we can identify two distinct electronic current transport mechanisms, hopping and interface-limited. The hopping transport dominates in the SrO-terminated structures, in which case the ideal LRS state would be the one with the uniform filament (\textbf{Fig.~\ref{fig:current}(b), left}). The interface-limited transport is being activated in the TiO$_2$-terminated structures, in which case the LRS state can be achieved for the vacancies located at the Pt end (\textbf{Fig.~\ref{fig:current}(a), right}). 
The spatial distributions of the electrical currents corresponding to these idealized LRS states are visualized in \textbf{Fig.~\ref{fig:current}(c)} in the form of isosurface values computed at a readout voltage of -1 V. In the SrO-terminated device, the current flow is concentrated near the vacancy sites across the oxide. On the other hand, in the TiO$_2$-terminated stack, the current flow is distributed more evenly throughout the oxide. 
 
The observed termination-dependence of the conduction mechanism in SrTiO$_3$ VCM cells thus originates from the relative alignment between the band edges of the oxide and the Fermi levels of the electrodes, which determine the Schottky barrier height. This alignment is strongly influenced by the bonding environment and, as a consequence, by the charge distribution at the metal-oxide interfaces. Hence, the two possible terminations of STO exhibit significantly different work functions \cite{STO_term_2} and Schottky barrier heights at the interface with the electrode \cite{STO_term_1}. The dominance of filamentary-type switching in SrO-terminated devices may explain why a higher performance and a larger forming voltage are typically observed in Sr-rich SrTiO$_3$ \cite{Z1, Z2}.

\subsection{Dynamic model of interface-type switching in SrTiO$_3$}

\begin{figure}[!t]
\centering\includegraphics[width=\columnwidth]{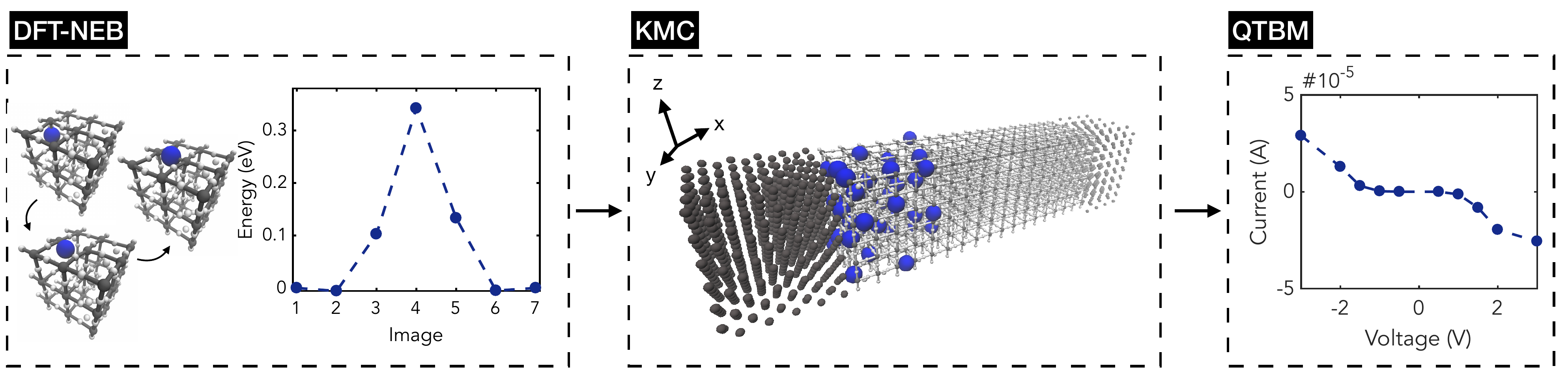}
\caption{Illustration of the simulation framework used for dynamic models of resistive switching. It consists of DFT-based nudged elastic band (NEB) calculations (left), kinetic Monte Carlo (KMC) simulations (middle), and I-V computations with the quantum transmitting boundary method (QTBM, right). The blue spheres represent oxygen vacancies. The data presented is for illustrative purposes only.}
\label{fig:method}
\end{figure}

As next step, we study the kinetics of interface-type switching in SrTiO$_3$ devices. For that purpose, we use a KMC solver capable of modeling the diffusion of oxygen vacancies under an applied bias as well as the influence of the device structure on its ``current vs. voltage'' characteristics (\textbf{Fig.~\ref{fig:method}}). We focus on TiO$_2$-terminated devices as they are more prone to exhibit interface-type switching. The KMC model employs a standard rejection-free algorithm \cite{KMC}, similar to what was previously used to study resistive switching in HfO$_2$\cite{ACS_NANO}. In essence, the KMC model selects and executes events based on their probabilities, which are defined by their activation energies (calculated with the NEB method) and by the electrostatic potential along the device resulting from the solution of Poisson's equation. More details about the model can be found in Ref.~\cite{ACS_NANO}. 

The main difference with respect to this original work comes from the treatment of the oxygen vacancy charges. In filament-containing structures, such as HfO$_2$-based VCMs, oxygen vacancies are assumed to be uncharged once they form clusters, which is the case within a conductive filament, and existing at a charge of +2 when isolated in the bulk oxide. In case of interface-type switching in SrTiO$_3$, a vacancy is reduced (and assigned a neutral charge state) if it is neighboring the electrode of a lower potential. This reduction is propagated to all vacancies located close to already-neutralized vacancies. A charge of +2 is assigned to all remaining vacancies. The distance used to identify neighboring oxygen vacancies in the percolation path method is set to 3.6 \AA, so that it includes nearest-neighbor vacancy pairs. An activation energy of 0.34 eV is calculated with the NEB method for vacancy diffusion. 

\begin{figure}
\centering
\includegraphics[width=400pt]{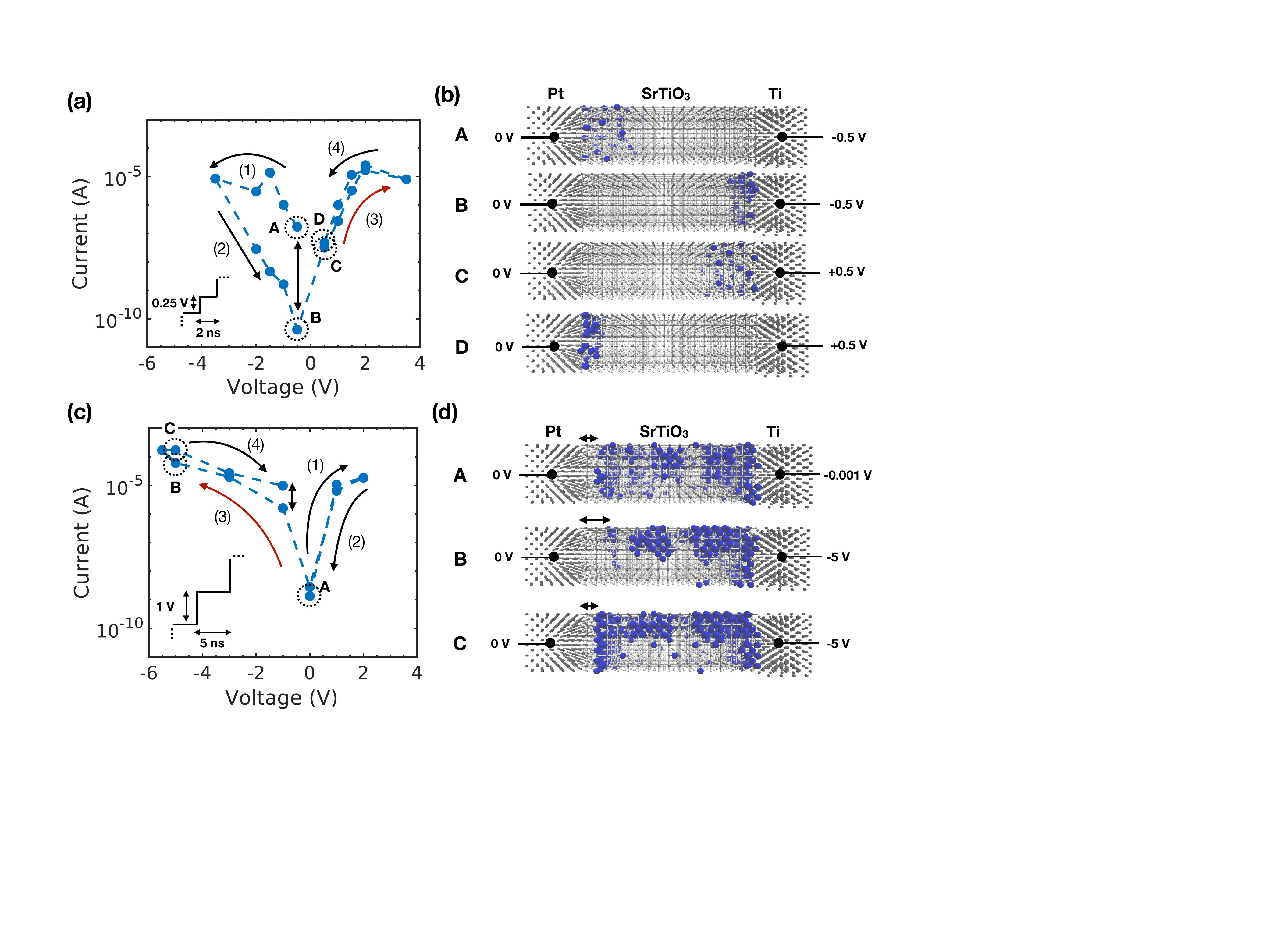}
\caption{(a) I-V characteristics of a TiO$_2$-terminated Pt-SrTiO$_3$-Ti stack exhibiting interface-type switching. The biasing scheme, shown as inset, corresponds to the voltage applied to the Ti electrode. The curved dashed arrows indicate the switching direction, while the vertical arrow the maximal dynamic range of around 10$^4$. 
(b) Atomic structures corresponding to data points in (a) denoted as A, B, C, and D. 
(c) I-V characteristics of a SrO-terminated Pt-SrTiO$_3$-Ti stack undergoing filamentary-type switching. The curved arrows indicate the switching directions, while the vertical arrow gives the maximal dynamic range of around 10. 
(d) Atomic structures corresponding to the data points denoted as A, B, and C in sub-plot (c).}
\label{fig8:iv}
\end{figure}

In the initial step of the KMC simulations, oxygen vacancies are randomly distributed at a concentration of 2.5\% across the SrTiO$_3$ layer before a 2 V pulse is set to the Ti electrode. This leads to a migration of the oxygen vacancies towards the Pt electrode, where they accumulate. After this initial step, the biasing scheme, shown as inset in \textbf{Fig.~\ref{fig8:iv}(a)}, is applied to the device. During the 0 V $\rightarrow$ -3.5 V sweep vacancies move towards the Ti electrode (configuration labeled as ``A'' in \textbf{Fig.~\ref{fig8:iv}(b)}): Since they are charged, they are accelerated by the external electric field before being reduced once they reach the Ti side. They then remain uncharged as long as a negative voltage is applied to the Ti electrode (-3.5 V $\rightarrow$ 0 V sweep, case  ``B''). Once the polarity is switched (the Pt electrode is now at a lower potential), the vacancies acquire back their positive charge and start migrating towards the Pt side (0 V $\rightarrow$ 3.5 V, structure ``C''). They stay there until the voltage goes back to 0 V, at the end of the switching cycle (configuration ``D''). 

Using the static model, we demonstrated that, under a negative applied voltage to the Ti electrode, the accumulation of vacancies at the Pt electrode leads to a higher current. This corresponds to the large conductance difference between points ``A'' and ``B'', between which the vacancies migrated away from the Pt electrode. The conductance difference between the two cases are not as pronounced under positive voltages (``C'' and ``D''), where current encounters the higher Schottky barrier at the Pt electrode. The induced imbalance in the I-V curve between the positive and negative polarities, also observed experimentally \cite{STO_7}, can thus be attributed to the asymmetry of the device stack: Only the Schottky barrier height at the Pt-STO electrode gets modulated.

Our simulated interface-type switching cycle is consistent with the switching direction (the counter-eightwise direction), shape of the I-V characteristics, and the assumed underlying physical mechanism \cite{STO_9, STO_11, STO_2, STO_3} of measurements on several fabricated SrTiO$_3$ devices. However, we have also found other experimental I-V curves of SrTiO$_3$ that share some features (such as the asymmetry and the nonlinearity of the I-V curve) with the results of our model, but exhibit a different switching direction \cite{STO_6, STO_3, STO_4}. We believe that in those cases, mechanisms other than vacancy migration are responsible for the switching. For example, they may be dominated by oxygen vacancy-ion pair generation at the Pt electrode or charge trapping/de-trapping at/from interface trap states. 

\subsection{Dynamic model of filamentary-type switching in SrTiO$_3$}

As suggested in previous reports \cite{STO_8, STO_10}, filamentary-type switching can also occur in SrTiO$_3$-based VCM cells. This requires an electro-forming of the device, during which a conductive filament is created. In more common VCM stacks, e.g., those relying on HfO$_2$, electro-forming is reported to take place at the active electrode, where oxygen ions are stored in an oxygen reservoir, leaving  oxygen vacancies behind in the oxide \cite{Dittmann2021, ACS_NANO}. Ti is known for its ability to partially oxidize and store oxygen ions, thus preventing recombination of generated defect pairs. On the other hand, experimental studies on SrTiO$_3$ devices clearly demonstrated the formation of oxygen vacancies at the Pt electrode, which is referred to as the active electrode. This implies that Pt and Ti electrodes, both commonly used contact metals in SrTiO$_3$ memristors, can be triggered for electro-forming and subsequent resistive switching, depending on the polarity and magnitude of the electroforming voltage. If SrTiO$_3$ is crystalline, the termination of this layer may impact the capacity to form a filament, rendering the functionality of these devices even more complex.

To shed light on filamentary-type switching in SrTiO$_3$ cells, we employ the same KMC model as in the previous section on interface-type switching. While the structural changes in the case of interface-type switching can be described by vacancy migration only, filamentary-type switching requires the inclusion of vacancy-ion pair generation and ion diffusion events. Through NEB calculations, we estimated the activation energy for vacancy-ion pair generation (an ion moving to an interstitial position leaving a vacancy behind) in bulk SrTiO$_3$ to be 1.70 eV, while the ion diffusion activation energy is equal to 0.51 eV. The activation energy for interface vacancy-ion pair generation depends on the electrode type due to the different bonding environment with SrTiO$_3$ and to a possible potential gradient caused by work function differences between the electrode and the oxide. Here, we calculated this activation energy for both electrodes (Pt and Ti) in case of SrO-terminated SrTiO$_3$ only, because filamentary switching dominates in this configuration, as discussed above. The resulting energies are 0.83 and 3.84 eV for the generation at the Pt and Ti electrode, respectively. Lastly, for filamentary-type switching, we employ the same charge condition for the oxygen vacancies as previously suggested for HfO$_2$ \cite{Lee2019}: If a vacancy has two or more neighboring vacancies within a radius of 3.6 \AA, its charge is set to 0, otherwise, it remains equal to +2. 

By combining KMC and quantum transport calculations, we can simulate the I-V curve corresponding to the filamentary-switching of the SrO-terminated SrTiO$_3$ layer (see \textbf{Fig.~\ref{fig8:iv}(c)}). As oxygen vacancy-ion pairs have a higher probability to be generated at the Pt-SrO interface, as demonstrated by the lower activation energy, we electro-form the filament by applying a positive voltage of 16 V to the Pt electrode (-16 V to the Ti electrode). The device then subjected to the biasing scheme shown in the inset of \textbf{Fig.~\ref{fig8:iv}(c)}. In our simulations, the bias is always applied to the Ti electrode, while the Pt electrode is grounded. Hence, the SET process occurs at negative voltages applied to the Ti electrodes, which are equivalent to positive voltages applied to the Pt contacts. The initially formed device (structure ``A'' in \textbf{Fig.~\ref{fig8:iv}(d)}) is reset at positive voltages. This RESET is caused by the recombination of oxygen vacancies and oxygen ions at the tip of the filament (structure ``B'') close to the Pt electrode, which leads to a decreased conductivity. The device stays in its HRS when the voltage is decreased to -5.5 V, at which point the SET process occurs. The latter is characterized by the generation of additional vacancy-ion pairs at the Pt electrode, thus shortening the tunneling gap and widening the filament (structure ``C''). The device stays in its LRS even when the voltage is dre-increased, which demonstrates its non-volatility. However, we note that the vacancies that are closest to the Pt electrode recombine with ions, creating a short tunneling gap even in the LRS state. Furthermore, the simulated switching is not as abrupt as commonly reported for filamentary devices. This may be attributed to the limited size of the device, similar to the effect reported for a Pt/TiO$_x$/HfO$_2$/Pt memristor\cite{Pi2019-ku}.

\section{Discussion}

Based on the aforementioned results, we find that the alignment between the Fermi levels of the contacts and the band edges of the STO layer crucially affect the underlying resistive switching mechanism. If the Fermi levels of the electrodes are close to the bottom of the conduction band STO, as in the TiO$_2$-terminated SrTiO$_3$ structures (\textbf{Fig.~\ref{fig:ldos_tio2}}), the modulation of the Schottky barrier height is the factor that mainly governs switching. On the other hand, if the Fermi levels of the contacts are farther from the oxide band edges, filamentary switching dominates. In that case the switching depends on the modulation of trap states in the STO band gap, as in SrO-terminated SrTiO$_3$ (\textbf{Fig.~\ref{fig:ldos_sro}}). In this case, switching (the closing/opening of a tunneling gap at the filament tip) is more likely to occur at the electrode that exhibits higher vacancy-ion pair generation rate, which is the Pt electrode in the devices considered here. Filamentary switching can be potentially observed if a positive voltage is applied to the counter electrode (here: Ti electrode), albeit with a higher SET voltage. Furthermore, we believe that filamentary switching may also be observed in TiO$_2$-terminated SrTiO$_3$ layers. However, a competition of two effects (modulation of the Schottky barrier vs. change in the tunneling gap) is expected to take place such that it becomes difficult to distinguish them. While our results provide insights into the origin of both interface- and filamentary-type switching in SrTiO$_3$, we cannot exclude that other mechanisms play an important role. For example, the generation of vacancies at the Pt electrode\cite{STO_6, STO_10} and/or charge trapping and de-trapping\cite{STO_5, STO_4, STO_7}. might also explain interface-type switching in SrTiO$_3$. The investigation of these mechanisms goes, however, beyond the scope of this work. 

\section{Conclusion}

We investigated the switching mechanism of SrTiO$_3$-based VCM cells using a fully atomistic model that combines DFT, KMC, and quantum transport calculations. We found that different terminations of the crystalline SrTiO$_3$ lead to different types of switching. While SrO-terminated structures are likely to exhibit only filamentary-type switching, their TiO$_2$-terminated counterparts tend to display interface-type switching through the modulation of the Schottky barrier height at the Pt end. Furthermore, we revealed that the modulation of the barrier is achieved through the migration and accumulation of oxygen vacancies at the electrode-oxide interface. Based on these findings, we calculated the I-V characteristics of a full switching cycle for a TiO$_2$-terminated (interface switching) and SrO-terminated (filamentary switching) Pt-SrTiO$_3$-Ti stack. Finally, we would like to emphasize that the structural properties of devices, such as the termination and the vacancy distribution, directly influence their resistive switching and can therefore be leveraged as design parameters, especially in the context of interface switching. 

\section{Supporting Information}

Details on device optimization, Hubbard parameter identification, KMC simulations, and the list of events used in KMC simulations with their activation energies.

\begin{acknowledgement}
The authors acknowledge funding from SNSF Sinergia (grant no.\;198612), the Werner Siemens Stiftung Center for Single Atom Electronics and Photonics, and the NSERC Postgraduate Scholarship. This work has received funding from the Swiss State Secretariat for Education, Research, and Innovation (SERI) under the SwissChips Initiative. Computational resources were provided by the Swiss National Supercomputing Center (CSCS) under projects s1119 and lp16. 
\end{acknowledgement}

\newpage 

\bibliography{main}

\end{document}